\documentclass[a4paper]{jpconf}
\usepackage{graphicx}
\begin{document}
\title{Existence of anticipatory, complete and lag synchronizations
in time-delay systems}

\author{D~V~Senthilkumar$^1$ and M~Lakshmanan$^2$}

\address{Centre for Nonlinear Dynamics, Department of Physics,
Bharathidasan University, Tiruchirapalli - 620 024, India}

\ead{$^1$skumar@cnld.bdu.ac.in}
\ead{$^2$lakshman@cnld.bdu.ac.in}
\begin{abstract}
Existence of different kinds of synchronizations, namely anticipatory, complete
and lag type synchronizations (both exact and approximate), are shown to be 
possible in time-delay coupled piecewise linear systems.  We deduce stability
condition for synchronization of such  unidirectionally coupled systems
following Krasovskii-Lyapunov theory.  Transition from anticipatory to lag
synchronization via complete synchronization as a function of coupling delay is
discussed.  The existence of exact synchronization is preceded by a region of
approximate synchronization from desynchronized state as a function of a system
parameter, whose value determines the stability condition for synchronization. 
The results are corroborated by the nature of similarity functions.  A new type
of oscillating synchronization that oscillates between anticipatory, complete
and lag synchronization, is identified as a consequence of delay time
modulation with suitable stability condition.
\end{abstract}

\section{Introduction}

Synchronization of coupled chaotic systems is a field of growing interest in
view of its potential applications in diverse area of research activities such
as secure communication, cryptography, controlling, long term prediction of
chaotic systems,
etc.\cite{aspmgr2001,mlkm1996,mlsr2003,shcg1993,kp1993,akkp1994}. Since the 
identification of chaotic synchronization different kinds of them have been
proposed in interacting chaotic systems both theoretically and experimentally:
complete synchronization \cite{hfty1983,lmptlc1990}, generalized
synchronization  \cite{nfrmms1995,rb1998}, phase synchronization
\cite{mgrasp1996,tyycl1997}, lag synchronization  \cite{mgrasp1997,mzgww2002}
and anticipatory synchronization\cite{huv2001,cm2001}.

One of the most important applications of chaos synchronization is secure
communication.  It has been shown that in secure communication based on simple
low dimensional chaotic systems with only one positive Lyapunov exponent, the
hidden message can be unmasked by dynamical reconstruction of the chaotic
signal using nonlinear dynamical forecasting methods or by using some simple
return maps \cite{kmsatp1998,gphac1995}.  One way to overcome this problem is
to consider chaos synchronization in higher dimensional systems having multiple
positive Lyapunov exponents, based on the consideration that increased
randomness and unpredictability of the hyperhcaotic signals will make it more
difficult to extract the masked message.  Recently, chaotic time-delay systems
have been suggested as good candidates for secure communication \cite{bmal1998}
as the time-delay systems exhibit the intriguing characteristics of increase in
the embedding dimension and the number of positive Lyapunov exponents with the
time-delay inspite of its small number of physical variables.  Therefore the 
study of chaos synchronization in time-delay systems is of considerable
practical significance.

Recently, we have shown that even a single scalar delay equation with piecewise
linear function can exhibit hyperchaotic behavior even for small values of
time-delay \cite{dvskml2005} and hence it is of interest to consider chaos
synchronization in such scalar piecewise linear delay differential
equations with appropriate coupling between them. For the present study, we will
consider chaos synchronization between two such time-delay systems with
unidirectional coupling and having two different time-delays: one  in the 
individual systems, namely, feedback delay $\tau_1$ and the other in the
coupling term $\tau_2$. We have arrived at the stability condition for
synchronization following Krasovskii-Lyapunov theory, which shows that the
stability condition is independent of the delay times, and demonstrate that
there exist transitions between three different kinds of synchronization,
namely, anticipatory, complete and lag synchronizations by simply tuning the
second time-delay parameter in the coupling, for a fixed set of other system
parameters satisfying appropriate stability condition. The results have been
corroborated by the nature of similarity functions.  To enhance the security,
delay time modulation is introduced at the feedback delay and the stability of
the synchronization manifold with delay time modulation is calculated.  We have
also identified a new type of oscillating synchronization as a consequence of
delay time modulation.

\section{Scalar piecewise linear delay system} 

\begin{figure}
\includegraphics[width=0.7\columnwidth]{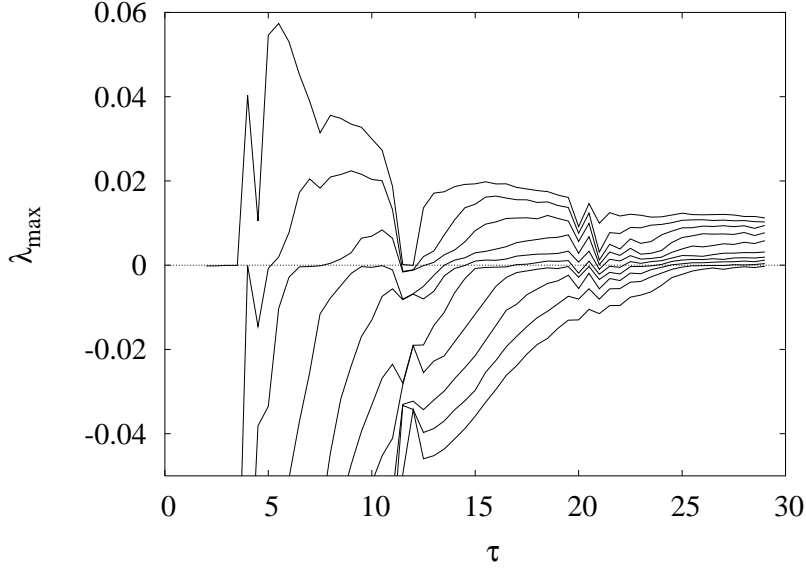}\hspace{1pc}%
\begin{minipage}[b]{10pc}\caption{\label{fig.one} The first ten maximal
Lyapunov exponents $\lambda_{max}$ of the scalar time-delay equation
(\ref{eqonea}) for the parameter values $a=1.0, b=1.2,$  $\tau\in(2,29)$.}
\end{minipage}
\end{figure}

We consider the following first order delay differential equation introduced by
Lu and He~\cite{hlzh1996} and discussed in detail in references
~\cite{dvskml2005,ptkm1998,dvskmlpre},
\begin{eqnarray}
\dot{x}(t)&=&-ax(t)+bf(x(t-\tau)),
\label{eqonea}
\end{eqnarray}
where $a$ and $b$ are parameters, $\tau$ is the constant time-delay and 
$f$ is an odd piecewise linear function defined by the relation
\begin{eqnarray}
f(x)=
\left\{
\begin{array}{cc}
0,&  x \leq -4/3  \\
            -1.5x-2,&  -4/3 < x \leq -0.8 \\
            x,&    -0.8 < x \leq 0.8 \\              
            -1.5x+2,&   0.8 < x \leq 4/3 \\
            0,&  x > 4/3 \\ 
         \end{array} \right.
\label{eqoneb}
\end{eqnarray}
This system exhibits hyperchaotic behavior for the parameter values $a=1.0,
b=1.2$ and $\tau=5$ and the hyperchaotic nature was confirmed by the existence
of multiple positive Lyapunov exponents.  The first ten maximal Lyapunov
exponents as a function of $\tau$ for the chosen parameter values is shown in
Fig.~\ref{fig.one}. In our present study, we have included three linear parts
for the function $f(x)$ in Eq.~\ref{eqoneb}.  One may also carry out the 
studies with two linear parts alone.  However, the dynamics in the three linear
part case is richer and so we confine our study to this case alone here.

\section{Stability condition for synchronization}

Now let us consider the following unidirectionally coupled drive $x_1(t)$ and 
response
$x_2(t)$ systems with two different time-delays $\tau_1$ and $\tau_2$ as
feedback and coupling time-delays, respectively,
\begin{equation}
\dot{x_1}(t)=-ax_1(t)+b_{1}f(x_1(t-\tau_{1})),
\label{eq.onea}
\end{equation}
\begin{equation}
\dot{x_2}(t)=-ax_2(t)+b_{2}f(x_2(t-\tau_{1}))+b_{3}f(x_1(t-\tau_{2})),
\label{eq.oneb}
\end{equation}
where $b_1, b_2$ and $b_3$ are constants, $a>0$, and $f(x)$ is of the same form
as in Eq.~(\ref{eqoneb}).

Now we can deduce the stability condition for synchronization of the two
time-delay systems Eqs.~(\ref{eq.onea}) and (\ref{eq.oneb}) in the presence
of the delay coupling  $b_{3}f(x_1(t-\tau_{2}))$.  
The time evolution of the difference system with the state variable
$\Delta=x_{1\tau}-x_2$ , where 
$x_{1\tau} = x_{1}(t-\tau), \tau=\tau_2-\tau_1$, can be  written for
small values of $\Delta$ by using the evolution Eqs.~(3) as
\begin{equation}
\dot{\Delta}=-a\Delta+(b_2+b_3-b_1)f(x_1(t-\tau_2))+b_2f^{\prime}
(x_1(t-\tau_2))\Delta_{\tau_1}, 
\end{equation}
In order to study the stability of the
synchronization manifold, we choose the parametric condition,
\begin{equation}
b_1=b_2+b_3,
\label{eq.paracon}  
\end{equation}
so that the evolution equation for the difference system $\Delta$ becomes
\begin{equation}
\dot{\Delta}=-a\Delta+b_{2}f^\prime(x_1(t-\tau_{2}))\Delta_{\tau_1}.
\label{eq.difsys}
\end{equation}
The synchronization manifold is locally attracting if the origin of this
equation is stable.  Following Krasovskii-Lyapunov functional approach 
\cite{nnk1963,kp1998},
 we define a positive
definite Lyapunov functional of the form
\begin{equation}
V(t)=\frac{1}{2}\Delta^2+\mu\int_{-\tau_1}^0\Delta^2(t+\theta)d\theta,
\end{equation}
where $\mu$  is an arbitrary positive parameter, $\mu>0$.  Note that $V(t)$
approaches zero as $\Delta \rightarrow 0$.

To estimate a sufficient condition for the stability of the solution $\Delta=0$,
we require the derivative of the functional $V(t)$ along the trajectory
of Eq.~(\ref{eq.difsys}),
\begin{equation}
\frac{dV}{dt}=-a\Delta^2+b_2f^{\prime}(x_1(t-\tau_2))\Delta
\Delta_{\tau_1}+\mu\Delta^2-\mu\Delta_{\tau_1}^2,
\end{equation}
to be negative.  After simple algebra, one can find the 
sufficient condition for asymptotic stability as
\begin{equation}
a>|b_2f^{\prime}(x_1(t-\tau_2))|
\label{eq.asystab}  
\end{equation}
along with the condition (\ref{eq.paracon}) on the parameters $b_1,b_2$ and $b_3$.

Now from the form of the piecewise linear function $f(x)$ given by Eq.~(2),
we have,
\begin{equation}
|f^{\prime}(x_1(t-\tau_2))|=
\left\{
\begin{array}{cc}
1.5,& 0.8\leq|x_1|\leq\frac{4}{3}\\
1.0,& |x_1|<0.8 \\
\end{array} \right.
\end{equation}
Consequently the stability condition
(\ref{eq.asystab}) becomes $a>1.5|b_2|>|b_2|$ along with the parametric
restriction $b_1=b_2+b_3$.

Thus one can take $a>|b_2|$ as a less stringent condition for (\ref{eq.asystab}) to
be valid, while
\begin{equation}
a>1.5|b_2|, 
\label{eq.four}
\end{equation}
as the most general condition specified by (\ref{eq.asystab}) for asymptotic
stability of the synchronized state $\Delta=0$.  The condition (\ref{eq.four})
indeed corresponds to the stability condition for exact anticipatory,
identical  as well as lag synchronizations for suitable values of the coupling
delay $\tau_2$. It may also be noted that the stability condition
(\ref{eq.four}) is independent of the both the delay parameters $\tau_1$ and
$\tau_2$.

\section{Anticipatory synchronization}

To start with, we demonstrate that there exists a region of anticipatory
synchronization when $\tau_2<\tau_1$ at the synchronization manifold.  For this
purpose, we have fixed the value of feedback delay $\tau_1$ at $\tau_1=0.25$
while the other parameters are fixed as $a=-0.16, b_1=0.2, b_2=0.1, b_3=0.1$
and the coupling delay $\tau_2$ is considered as a control parameter.  For 
all  values of $\tau_2<\tau_1$, we find from the evolution equation
(\ref{eq.difsys}), the difference variable $\Delta=x_1(t-\tau)-x_2(t)$ with
$\tau=\tau_2-\tau_1<0$ approaches zero asymptotically, so that 
$x_1(t+\tau)=x_2(t)$, thereby demonstrating the existence of anticipatory
synchronization with anticipating time as $\tau$.  The existence of
anticipatory synchronization is characterized by the nature of similarity
functions for  different values of $b_2$, the parameter whose value determines
the stability of the synchronization manifold as may be seen from 
Eqn.~(\ref{eq.four}).

Now we use the notion of similarity function to characterize anticipatory
synchronization which was originally introduced by Rosenblum \emph{et al.} \cite{mgrasp1996} to
characterize lag synchronization.  Similarity function is defined as a
time-averaged difference between the variables $x_1$ and $x_2$ (with mean
values being subtracted) taken with the positive time shift $\tau$ for lag
synchronization  and negative time shift $-\tau$ for anticipatory
synchronization,
\begin{eqnarray}
S_a^2(\tau)=\frac{\langle[x_2(t-\tau)-x_1(t)]^2\rangle}
{[\langle x_{1}^2(t)\rangle\langle x_{2}^2(t)\rangle]^{1/2}},
S_l^2(\tau)=\frac{\langle[x_2(t+\tau)-x_1(t)]^2\rangle}
{[\langle x_{1}^2(t)\rangle\langle x_{2}^2(t)\rangle]^{1/2}},
\label{anti:sim}
\end{eqnarray}
where $\langle x \rangle$ means time average over the variable $x$. If the
minimum value of $S_a(\tau)$ reaches zero, that is $S_a(\tau)=0$, then there
exists a time shift $-\tau$ between the two signals $x_1(t)$ and $x_2(t)$ such
that $x_2(t)=x_1(t+\tau)$, demonstrating the existence of anticipatory
synchronization between the drive $x_1$ and the response $x_2$ signals.
\begin{figure}
\includegraphics[width=0.6\columnwidth]{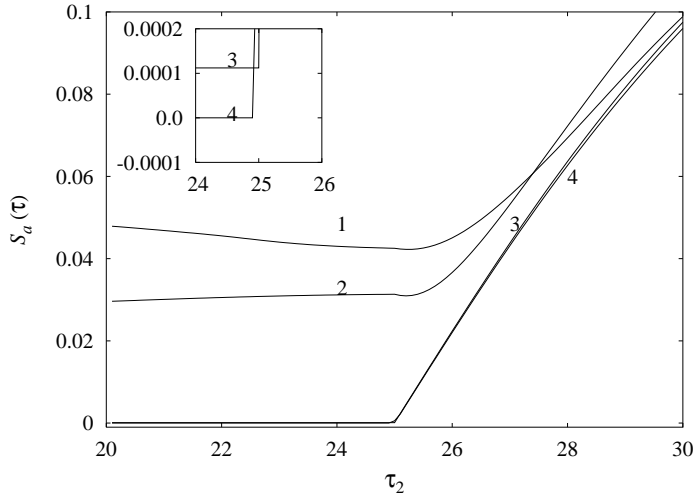}\hspace{1pc}%
\begin{minipage}[b]{14pc}\caption{\label{fig.asim}Similarity function
$S_a(\tau)$ for different values of $b_2$, the other system parameters are
$a=0.16, b_1=0.2$ and  $\tau_1=25.0$. (Curve~1: $b_2=0.18, b_3=0.02$, Curve~2:
$b_2=0.16, b_3=0.04$,   Curve~3: $b_2=0.15, b_3=0.05$ and Curve~4: $b_2=0.1,
b_3=0.1$).}
\end{minipage}
\end{figure}
Fig.~\ref{fig.asim} shows the similarity function $S_a(\tau)$ as a  function of
the coupling delay $\tau_2$ for four different values of $b_2$, the parameter
whose value determines the stability condition given by Eq.~(\ref{eq.four}),
while satisfying the parametric condition $b_1=b_2+b_3$.  Curves 1 and 2 are
plotted for the values of $b_2 = 0.18 (>a=0.16>a/1.5)$ and $b_2 = 0.16
(=a>a/1.5)$, respectively, where the minimum values of $S_a(\tau)$ is found to
be greater than zero,  indicating that there is no exact time shift between the
two signals $x_1(t)$ and $x_2(t)$. Note that in the both cases the stringent
stability condition (\ref{eq.four}) and the less stringent condition $a>|b_2|$
are violated. Curve 3 corresponds to the value of $b_2=0.15$ (
$<a>a/1.5$), where the minimum value of $S_a(\tau)$ is
almost zero, but not exactly zero (as may be seen in the inset of
Fig.~\ref{fig.asim}, inwhich case only the less stringent condition $a>|b_2|$
is satisfied while the stringent stability condition is ruled out), indicating an approximate anticipatory synchronization
$x_1(t)\approx x_2(t-\tau)$. On the other hand the curve 4 is plotted for the
value of $b_2 = 0.1 (<a/1.5)$, satisfying the  general stability criterion,
Eq.~(\ref{eq.four}).  It shows that the minimum of $S_a(\tau) = 0$, thereby
indicating that there exists an exact time shift between the two signals
demonstrating anticipatory synchronization.  The  anticipating time is found to
be equal to the difference between the coupling and feedback delay times, that
is, $\tau=\tau_2-\tau_1$. Note that $S_a(\tau) = 0$ for all values of $\tau_2
<\tau_1$, indicating anticipatory synchronization for a range of delay
coupling.  A further significance is that the anticipating time  $\tau =
\tau_2-\tau_1$ is an adjustable quantity as long as $\tau_2 <\tau_1$, which
can be tuned suitably to satisfy experimental situations. Thus anticipating of
any future chaotic state can be made possible. We also note that the above 
choice of parameters (Curve 4) in Fig.~\ref{fig.asim} is only a specific
example exhibiting exact anticipatory synchronization. Any other
choice of parameters satisfying the condition $b_1=b_2+b_3$ subject to
$a>1.5|b_2|$ is good enough to obtain the above phenomenon.

The existence of
exact synchronization is preceded by a region of approximate synchronization from
desynchronized state as a function of the parameter $b_2$, whose value
determines the stability condition for synchronization.
We have also found \cite{dvskmlpre}that the existence of approximate
synchronization from desynchronized state is characterized by transition from
on-off intermittency to periodic structure in probability distribution of
laminar phase as shown by Zhan \emph{et al}\cite{mzgww2002}.

\section{Complete synchronization}

Now, we show that when the value of coupling delay $\tau_2$ equals the value of
feedback delay $\tau_1$, that is, $\tau_2=\tau_1$, there exists complete
synchronization.  We find from the evolution equation (\ref{eq.difsys}), the
difference variable $\Delta=x_1(t-\tau)-x_2(t)$ with $\tau=\tau_2-\tau_1=0$
approaches zero asymptotically, so that $x_1(t)=x_2(t)$, thereby demonstrating
the existence of  complete synchronization between the drive $x_1(t)$ and the
response $x_2(t)$.  Here also, we have identified that the emergence of
approximate complete synchronization is associated with the transition form
on-off intermittency to periodic structure in the laminar phase distribution as
a function of the parameter $b_2$.
\begin{figure}
\includegraphics[width=0.6\columnwidth]{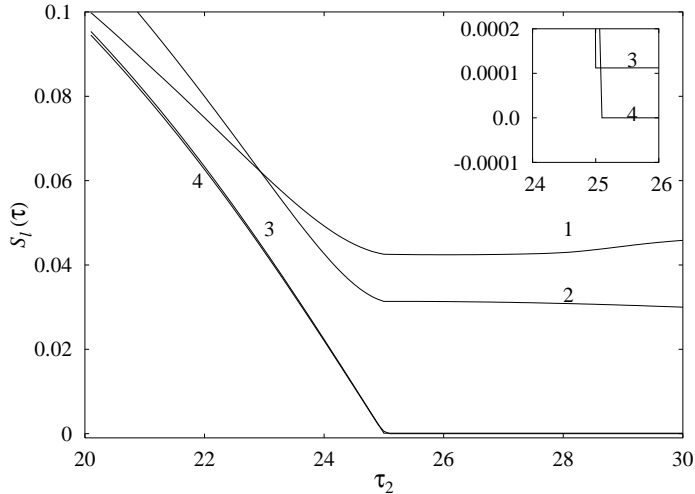}\hspace{1pc}%
\begin{minipage}[b]{14pc}\caption{\label{lsim}Similarity function $S_l(\tau)$
for different values of $b_2$, the other system parameters are $a=0.16,
b_1=0.2$ and  $\tau_1=25.0$. (Curve~1: $b_2=0.18, b_3=0.02$, Curve~2:
$b_2=0.16, b_3=0.04$  Curve~3: $b_2=0.15, b_3=0.05$ and Curve~4: $b_2=0.1,
b_3=0.1$).}
\end{minipage}
\end{figure}

\section{Lag synchronization}

Next, we show that for the value of coupling delay $\tau_2$ greater than
feedback delay $\tau_1$, the system exhibits lag synchronization.  For
$\tau_2>\tau_1$,  we find from the evolution equation (\ref{eq.difsys}), the
difference variable $\Delta=x_1(t-\tau)-x_2(t)$ with $\tau=\tau_2-\tau_1>0$
approaches zero asymptotically, so that  $x_1(t-\tau)=x_2(t)$, thereby
demonstrating the existence of lag synchronization with lag time equal to
$\tau=\tau_2-\tau_1$.  The existence of lag synchronization is  characterized
by the similarity function $S_l(t)$ defined earlier in  Eqn.~(\ref{anti:sim}).
Fig.~\ref{lsim} shows the similarity function $S_l(\tau)$ Vs coupling delay 
$\tau_2$ for four different values of $b_2$.  Curves 1 and 2 show the
similarity function $S_l(\tau)$ for the values of $b_2 = 0.18$ and $0.16$,
respectively.  The minimum of similarity function $S_l(\tau)$ occurs for 
values of $S_l(\tau) > 0$, where the minimum does not occur at zero and hence
there is a lack of  exact lag time  between the drive and response signals
indicating asynchronization. Curve 3 corresponds to the value of $b_2=0.15$
(which is less than $a$ but greater than $a/1.5$), where the minimum values of
$S_l(\tau)$ is almost zero, but not exactly zero  (as may be seen in the inset
of Fig.~\ref{lsim}), so that $x_1(t)\approx x_2(t+\tau)$.   However for the
value of $b_3 = 0.1$, for which the general condition~(\ref{eq.four}) is
satisfied, the minimum of similarity function  becomes exactly zero (Curve 4)
indicating that there is an exact time shift (Fig.~\ref{lsim}) between drive
and  response signals $x_1(t)$ and $x_2(t)$,  respectively, confirming the
occurrence of lag synchronization.  Again the above  choice of parameters
(Curve 4) in Fig.~\ref{lsim} is only a specific example exhibiting exact lag
synchronization. Any other choice of parameters satisfying the condition
(\ref{eq.paracon}) subject to $a>1.5|b_2|$ is good enough to obtain the above
phenomenon.

We have also confirmed that as in the case of anticipatory synchronization, when
the parameter $b_2$ varies, the onset of exact lag synchronization is preceded
by a region of approximate lag synchronization, which is  characterized by a
transition from on-off intermittency of the desynchronized state to a periodic
structure in the laminar phase distribution.

\section{Stability condition for synchronization with delay time modulation}

To improve the security of the above type of synchronized systems, we have
extended our studies by choosing time-delay as a function of time.  Recently,
the concept of delay time modulation was introduced by Kye \emph{et al} 
\cite{whkmc2004,whkmcpre} and they have shown that reconstruction of phase space of the
time-delay systems is hardly possible as a consequence of delay time modulation.

Now, we will introduce the delay time modulation in the systems (\ref{eq.onea})
and (\ref{eq.oneb}) at the feedback delays $\tau_1(t)$ in the form
\cite{emskas2005} 
\begin{eqnarray}
\tau_1(t)&=&\tau_0+\tau_a\sin(\omega t),
\label{eqonec}
\end{eqnarray}
where $\tau_0$ is the zero frequency component, $\tau_a$ is the amplitude and 
$\omega/2\pi$ is the frequency of the modulation. Here the coupling delay 
$\tau_2$ is kept  constant. 

The stability condition for synchronization of the systems (\ref{eq.onea}) and
(\ref{eq.oneb}) with delay time modulation can be obtained as (by following the
same procedure as for the case of constant feedback delay given in section 3),
\begin{equation}
a>\left|\frac{b_2f^{\prime}(x_1(t-\tau_2))}{\sqrt{(1-\tau_1^{\prime})}}\right|
\label{eq.asystab1}  
\end{equation}
along with the condition (\ref{eq.paracon}) on the parameters $b_1,b_2$ and $b_3$.
Once again as in the previous case, from the form of piecewise function $f(x)$,
one can take $a>\left|\frac{b_2}
{\sqrt{(1-\tau_1^{\prime})}}\right|$ as a less stringent condition for 
(\ref{eq.asystab1}) to be valid, while
\begin{equation}
a>1.5\left|\frac{b_2}{\sqrt{(1-\tau_1^{\prime})}}\right| 
\label{eq.four1}
\end{equation}
as the most general condition specified by (\ref{eq.asystab1}) for asymptotic
stability of the synchronized state.

\begin{figure}
\begin{center}
\includegraphics[width=1.0\columnwidth]{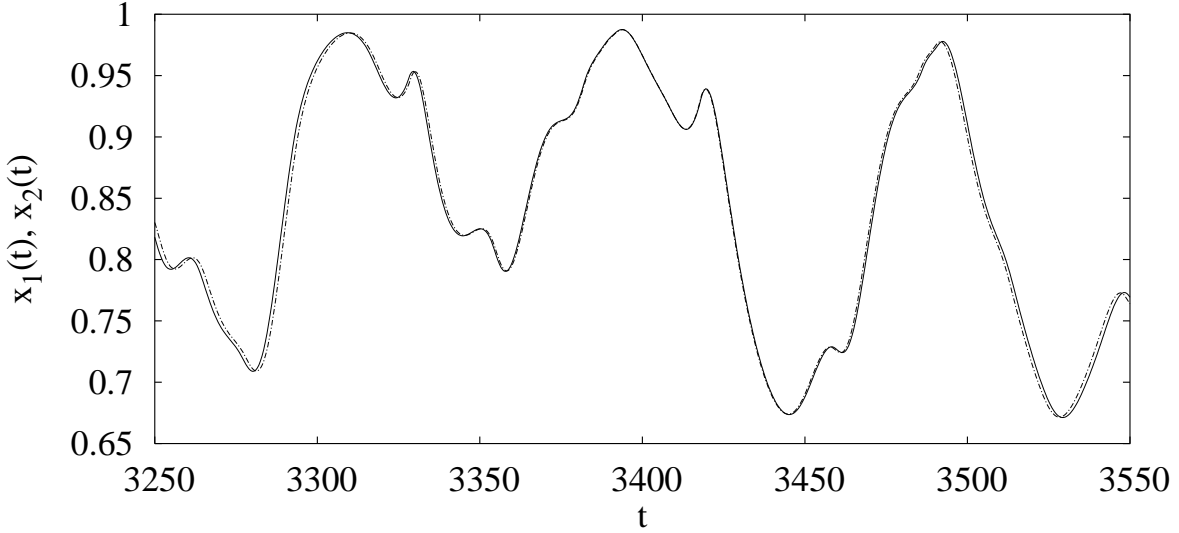}
\end{center}
\caption{\label{osc}Oscillating synchronization from lag  to anticipatory
synchronization via complete synchronization for $\tau_0=100, \tau_a=90$ and
$\omega=10^{-4}$. Drive $x_1(t)$ is represented by $\full$ and response
$x_2(t)$ by $\chain$.}
\end{figure}

\section{Oscillatory synchronization as a consequence of delay time modulation}

\begin{figure}
\begin{center}
\includegraphics[width=1.0\columnwidth]{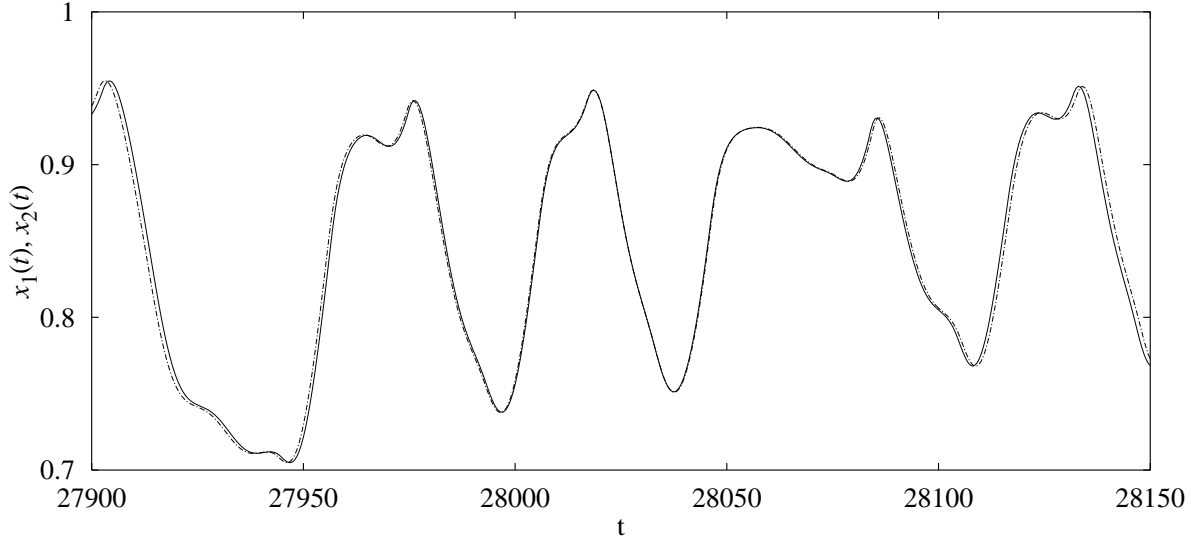}
\end{center}
\caption{\label{osc2}Oscillating synchronization from anticipatory  to lag
synchronization via complete synchronization for $\tau_0=100, \tau_a=90$ and
$\omega=10^{-4}$. Drive $x_1(t)$ is represented by $\full$ and response
$x_2(t)$ by $\chain$.}
\end{figure}

We have fixed the value of $\tau_0=100, \tau_a=90$ and $\omega=10^{-4}$ with
the same value of $a$ and $b$ as previously studied.  For the value of
$b_2=0.01$ the general stability condition (\ref{eq.asystab1}) is satisfied.
For the chosen values of $\tau_0$ and $\tau_a$, one can find that $\tau_1$
oscillates between  ($\tau_1(t)=\tau_0+\tau_a\sin(\omega t)=100 \pm 90$) 10
and 190.  Hence for any value of coupling delay $\tau_2$ less than 10,
$\tau_2$ is always less than $\tau_1(t)$, one can find from the evolution
equation (\ref{eq.difsys}) the difference variable
$\Delta=x_1(t-\tau(t))-x_2(t)$ with $\tau(t)=\tau_2-\tau_1(t)$ approaches zero
asymptotically, so that  $x_1(t+\tau(t))=x_2(t)$, implying the existence of
anticipatory synchronization with the time dependent anticipating time
$\tau(t)=\tau_2-\tau_1(t)$.  Similarly for any value of coupling delay
$\tau_2$ greater than 190, that is for $\tau_2$  always greater than
$\tau_1(t)$, we have $x_1(t-\tau(t))=x_2(t)$, demonstrating the existence of
lag synchronization.  On the other hand, when the coupling delay $\tau_2$ is
chosen in between 10 and 190, one is able to find that the feedback delay
$\tau_1(t)$ is less than $\tau_2$ for some time  (inwhich case
$\tau(t)=\tau_2-\tau_1(t)>0$, so that there exists lag synchronization
$x_1(t-\tau(t))=x_2(t)$) and as $\tau_1(t)$ increases eventually, it
approaches $x_2(t)$  at a certain time, where $\tau(t)=\tau_2-\tau_1(t)=0$, so
that $x_1(t)=x_2(t)$, that is there exists complete synchronization for a
specific value of time.  As $\tau_1(t)$ exceeds the value of $\tau_2$, now
$\tau(t)$ becomes negative, $\tau(t)=\tau_2-\tau_1(t)<0$ with
$x_1(t+\tau(t))=x_2(t))$, there exists anticipatory synchronization. Therefore
as time evolves there is oscillation between lag, complete and anticipatory
synchronization with time dependent anticipating and lag times. Fig.~\ref{osc}
shows the evolution of the drive $x_1(t)$ and the response  $x_2(t)$ at the
transition between lag to anticipatory synchronization via complete
synchronization, whereas  Fig.~\ref{osc2} shows the evolution of the drive
$x_1(t)$ and the response  $x_2(t)$ at the next transition between
anticipatory to lag via complete synchronization. Thus as a consequence of
delay time modulation there exists a new type of oscillating synchronization
that oscillates between lag, complete and anticipatory synchronization with
varying anticipating and lag times.

\section{Summary and conclusion}

In summary, we have obtained the stability condition for synchronization in a
system of two coupled piecewise linear differential equation following
Krasovskii-Lyapunov functional approach for both the case of constant and time
dependent delay.  We have demonstrated that there exists regimes of
anticipatory, complete and lag synchronizations with coupling delay $\tau_2$
as control parameter for constant delay case with anticipating and lag time as
$\tau=\tau_2-\tau_1$.  The exact synchronization regimes are preceded by a
regime of approximate synchronization from the desynchronized state as a
function of parameter $b_2$, which determines the stability condition.  The
existence of anticipatory and lag synchronizations are characterized by the
similarity functions.  In the case of time dependent delay, we have
demonstrated that there exists regimes of anticipatory and lag synchronization
with time dependent anticipating and lag times for suitable values of coupling
delay $\tau_2$.  We have also shown the existence of new type of oscillating
synchronization that oscillates between lag, complete and anticipatory
synchronizations as a consequence of delay time modulation, when the value of
coupling delay $\tau_2$ is chosen in between $\tau_1=\tau_0-\tau_a$ and
$\tau_1=\tau_0+\tau_a$ with varying anticipating and lag times.  We believe
that this new type of oscillating synchronization with  time dependent
anticipating and lag times  will enhance the security of secure communication
and now we are implementing this experimentally through electronic circuits.

\ack
This work has been supported by a Department of Science and Technology,
Government of India sponsored research project.

\end{document}